\documentclass{article}
\usepackage[]{epsfig}
\usepackage{epsf}
\usepackage[centertags]{amsmath}
\usepackage{amsfonts}
\newcommand{\be}{\begin{equation}}
\newcommand{\ee}{\end{equation}}
\newcommand{\bea}{\begin{eqnarray}}
\newcommand{\eea}{\end{eqnarray}}
\newcommand{\ba}{\begin{eqnarray}}
\newcommand{\ea}{\end{eqnarray}}

\newcommand{\pu}{\bar\theta^{\dot \alpha}}

\begin{document}
%%%%%%%%%%%%%%%%%%%%%%%%%%%%%%%%%%%%%%%%%%%%%%%%%%%%%%%%%%%%%%%%%%%
\title{Structure of the Vacuum in Deformed Supersymmetric Chiral Models}
\author{P.~Fern\'andez$^a$, E.~F.~Moreno$^b$,
F.~A.~Schaposnik$^a$\thanks{Associated with CICBA}
\\
\\
{\normalsize $^a\!$\it Departamento de F\'\i sica, Universidad Nacional de
La Plata}\\ {\normalsize\it C.C. 67, 1900 La Plata,
Argentina}
\\
{\normalsize $^b\!$\it Department of Physics,West Virginia University}\\
{\normalsize\it Morgantown, West Virginia 26506-6315, U.S.A.}
}
%\date{\hfill}
\maketitle
%===================================================================
\begin{abstract}
We analyze the vacuum structure of ${\cal N} = 1/2$ chiral
supersymmetric theories in deformed superspace. In particular
we study O'Raifeartaigh models with $C$-deformed
superpotentials and canonical and non-canonical deformed
K\"ahler potentials. We find conditions under which the vacuum
configurations are affected by the deformations.
\end{abstract}
\section{Introduction}

A $C$-deformation of the ${\cal N}=1$ superalgebra
corresponding to nonanticommutative Grassmann coordinates
$\theta^\alpha$ has been shown to arise in string theory in a
graviphoton background \cite{deBoer}-\cite{Ooguri}. Prompted by
this result, nonanticommutative versions of supersymmetric
(SUSY) Yang-Mills theory and Wess-Zumino model have been
formulated \cite{Sei1}-\cite{Araki} and their renormalizability
established \cite{Lunin}-\cite{Britto1}. The deformation
preserves the notion of chirality but only half of the ${\cal
N} =1$ supersymmetry is preserved as the supercharges
$Q_\alpha$, the generators of $\theta_\alpha$ translations, are
conserved while the $\bar Q_{\dot \alpha}$ are broken
explicitly.

In order to  analyze  the vacuum structure of undeformed SUSY
chiral models we study the effective potential $V$ for scalar
fields since its critical points correspond to the possible
vacua. Hermiticity of the original theory guarantees that the
resulting potential is positive definite so that the vanishing
of $V$ implies the existence of a supersymmetric vacuum. But in
$C$-deformed SUSY theories hermiticity is lost, $V$ is not
positive definite and the analysis of the critical points
should be done at the quantum level using saddle point or
steepest descent methods.

The issue of spontaneous supersymmetry breaking in
O'Raifeartaigh models \cite{O'R} has recently received much
attention after the discovery of meta-stable SUSY breaking
vacua in ${\cal N} =1$ SQCD that can be seen, in the low-energy
effective theory, as vacua of an O'Raifeartaigh-type model
\cite{Sei0}-\cite{Ray}. In connection with this phenomenon, it
is the purpose of this work to analyze the structure of the
vacuum for $C$-deformed O'Raifeartaigh-like models, discussing
in particular the possibility of spontaneous breaking of the
surviving supersymmetry.

As explained in \cite{Britto} a ${\cal N} = 1/2$ supersymmetric
vacuum requires both the $|{\rm vac}\rangle$ state and its dual
$ \langle{\rm vac}|$ to be annihilated by $Q_\alpha$. This is
connected to the fact that  the vacuum energy of such state,
$\langle{\rm vac}|E|{\rm vac}\rangle$, vanishes even if the
energy associated with the non-Hermitian deformed Lagrangian is
in general complex-valued. Hence, the analysis  of the zeroes
of the scalar potential still provides information about
symmetry breaking in deformed models and this is the route we
will follow in this investigation.

A discussion of the scalar potential for certain SUSY deformed
models has been already presented in
refs.\cite{ketov}-\cite{Chu} for deformed Wess-Zumino and sigma
models (with canonical K\"ahler potentials). Here we will
consider O'Raifeartaigh models with more general deformed
superpotentials and we  will also discuss the case of deformed
non-canonical K\"ahler potentials. The plan of the paper is the
following: In section 2 we establish our conventions for
nonanticommutative superspace and present general deformed
models containing chiral superfields. In section 3 we analyze
the vacuum structure of rather general deformed
O'Raifeartaigh-like models in which the K\"ahler potential is
kept canonical, and in section 4 a similar analysis of deformed
models with non-canonical K\"ahler potential. We summarize and
discuss our results in section 5.

%%%%%%%%%%%%%%%%%%%%%%%%%%%%%%%%%%%%%%%%%%%%%%%%%%%%%%%%%%%%%%%%%%%%%%%%%%
%%%%%%%%%%%%%%%%%%%%%%%%%%%%%%%%%%%%%%%%%%%%%%%%%%%%%%%%%%%%%%%%%%%%%%%%%%
%%%%%%%%%%%%%%%%%%%%%%%%%%%%%%%%%%%%%%%%%%%%%%%%%%%%%%%%%%%%%%%%%%%%%%%%%%
%%%%%%%%%%%%%%%%%%%%%%%%%%%%%%%%%%%%%%%%%%%%%%%%%%%%%%%%%%%%%%%%%%%%%%%%%%

\section{Non(anti)commutative superspace and chiral models}

%%%%%%%%%%%%%%%%%%%%%%%%%%%%%%%%%%%%%%%%%%%%%%%%%%%%%%%%%%%%%%%%%%%%%%%%%%
%%%%%%%%%%%%%%%%%%%%%%%%%%%%%%%%%%%%%%%%%%%%%%%%%%%%%%%%%%%%%%%%%%%%%%%%%%
\subsection{The setting}

We consider the deformation of 4 dimensional Euclidean ${\cal
N} = 1$ superspace parametrized by superspace bosonic
coordinates $x^\mu$ and chiral and anti-chiral fermionic
coordinates $\theta^\alpha,\bar\theta^{\dot \alpha}$ as
proposed in \cite{Sei1}
\be \{\theta^\alpha, \theta ^\beta\} = C^{\alpha \beta}
\label{uno}
\ee
\be \{\pu,\bar\theta^{\dot \beta}\} = 0 \; , \;\;\; \;\;\;
\{\theta^\alpha,\bar\theta^{\dot \beta}\} = 0 \label{dos0} \ee
Here $C^{\alpha \beta}$ are constant  elements of a symmetric
matrix. Defining chiral and anti-chiral coordinates according to
\begin{align}
y^\mu &= x^\mu +i \theta \sigma^{\mu} \bar\theta\\
\bar y^\mu &= y^\mu -2i \theta \sigma^{\mu} \bar\theta
\label{antichi}
\end{align}
we impose
\be [y^\mu,y^\nu] = [y^\mu, \theta^\alpha] = [y^\mu,\pu] = 0
\label{dos} \ee
and obtain, as a consequence of (\ref{uno})-(\ref{dos}),
\be
[\bar y^\mu,\bar y^\nu] = 4\bar\theta\bar\theta
\mathbf{C}^{\,\mu\nu}.
 \label{tres}
\ee
where $\mathbf{C}^{\,\mu\nu} = C^{\alpha\beta}
(\sigma^{\mu\nu})_{\alpha\beta}$ is antisymmetric and
antiselfdual.

The non(anti)commutative field theory in such a deformed
superspace can be defined in terms of superfields that are
multiplied according to the following  Moyal product
\cite{Sei1}
\be \Phi(y, \theta, \bar \theta) * \Psi (y, \theta, \bar \theta) =
\Phi (y, \theta, \bar \theta) \exp\left(-\frac{C^{\alpha\beta}}{2}
\frac{\overleftarrow{\partial}}{\partial \theta^\alpha} \frac{
\overrightarrow{\partial}}{\partial \theta^\beta} \right) \Psi(y,
\theta, \bar \theta) \label{moyprod} \ee
Supercharges and covariant derivatives in chiral coordinates take
the form \be Q_\alpha = \frac{\partial}{\partial \theta^\alpha} \; ,
\;\;\;\;\; \;\;\;\;\; \;\;\;\;\; \;\;\;\;\; {\bar Q}_{\dot \alpha} =
- \frac{\partial}{\partial {\bar\theta}^{\dot\alpha}} + 2i
\theta^\alpha \sigma^\mu_{\alpha\dot\alpha} \frac{\partial}{\partial
y^\mu}, \ee \be {D}_{ \alpha} = \frac{\partial}{\partial
{\theta}^{\alpha}} + 2i \sigma^\mu_{\alpha\dot\alpha}{\bar
\theta}^{\dot\alpha} \frac{\partial}{\partial y^\mu}  \; , \;\;\;
\;\; \;\;\;\;\; \;\;\;\;\; \;\;\;\;\; {\bar D}_{\dot \alpha} = -
\frac{\partial}{\partial {\bar\theta}^{\dot \alpha}} \ee
The $D$-$D$ algebra is not modified by the deformation
(\ref{uno}) as it also happens for the  $Q$-$D$ and $\bar
Q$-$D$ algebra. Concerning the supercharge algebra, it is
modified according to
\ba
 \{{\bar Q}_{\dot \alpha},Q_\alpha \} &=&
 2i \sigma^\mu_{\alpha \dot \alpha} \frac{\partial}{\partial y^\mu} =
2  \sigma^\mu_{\alpha \dot \alpha} P_\mu
  \label{nodef}\\
 \{Q_\alpha,Q_\beta\} &=& 0\\
 \{{\bar Q}_{\dot \alpha}, {\bar Q}_{\dot \beta} \} &=& -4
 C^{\alpha\beta} \sigma^\mu_{\alpha \dot \alpha}\sigma^\nu_{\beta
 \dot \beta} \frac{\partial^2}{\partial y^\mu \partial y^\nu}
 = 4
 C^{\alpha\beta} \sigma^\mu_{\alpha \dot \alpha}\sigma^\nu_{\beta
 \dot \beta} P_\mu P_\nu
\ea
Then, only the subalgebra generated by $Q_\alpha$ is still preserved
and this defines the chiral ${\cal N} = 1/2$ supersymmetry algebra
\cite{Sei1}.

%%%%%%%%%%%%%%%%%%%%%%%%%%%%%%%%%%%%%%%%%%%%%%%%%%%%%%%%%%%%%%%%%%%%%%%%%%
%%%%%%%%%%%%%%%%%%%%%%%%%%%%%%%%%%%%%%%%%%%%%%%%%%%%%%%%%%%%%%%%%%%%%%%%%%
\subsection{Chiral models}

In this work we will discuss models containing chiral
superfields. In deformed superspace, a chiral superfield $\Phi$
satisfying $\bar D_{\dot \alpha} \Phi = 0$ can be written, as
usual, in the form
\be \Phi(y,\theta) = \phi(y) + \sqrt 2\, \theta \psi(y) +
\theta\theta F(y) \label{chiral13} \ee
Analogously we can define antichiral superfields satisfying
\be D_{\alpha} {\bar \Phi}=0 \ee
which only depend on ${\bar \theta}$ and ${\bar y}^{\mu}$.

A general action in terms of chiral and antichiral superfields
takes the form
\be \label{NLSM} S\left[\Phi,\bar\Phi\right]=\int d^4y
\left[\int d^2 \theta d^2\bar\theta \,
K_*\left(\Phi^i,\bar\Phi^{\bar j}\right) +\int d^2\theta \,
W_*\left(\Phi^i\right)+\int d^2 \bar\theta \, \bar
W_*\left(\bar\Phi^{\bar j}\right)\right] \ee
Here we call  $K_*$,  $W_*$ the K\"ahler and superpotential
functionals with superfields multiplied using the Moyal
product. A very useful formula for handling these quantities
has been derived in  \cite{AG}-\cite{ketov2}. For example,
given the superpotential $W_*(\Phi)$,  we can define a
``diffuse superpotential''
\be \widetilde{\mathcal{W}}\left(\phi_i,F_i\right) =
\int^1_{-1}d\xi \; {\cal W}\left(\phi_i+\xi cF_i\right)
\label{wtilde} \ee
where fields  $\phi_i$ are multiplied in the r.h.s. with the
ordinary product and we have written $ c=\sqrt{-\det C}$. As
pointed out in \cite{AG}, non(anti)commutativity induces
certain fuzziness controlled by auxiliary fields $F_i$.

Using eq.(\ref{wtilde}), we can prove that, in terms of
component fields, the scalar potential can be written
\begin{align}
\label{Vescalar}
V_{scalar}\left(\phi_i,\bar\phi_{\bar i}\right)=
\left.\frac{1}{2}F_i\widetilde{W},_i\right\vert_{F_i =
F_i\left(\phi,\bar\phi\right)}
\end{align}
with all products being ordinary products. Analogously, we can
define, starting from the K\"ahler potential, the following
diffuse quantities \cite{ketov2}
\begin{align}
\label{Zgral}
Z\left(\phi,\bar\phi,F\right)=&\int^1_{-1}d\xi
K\left(\phi_i+\xi cF_i,\bar\phi_{\bar j}\right)\\
Y\left(\phi,\bar\phi,F,\bar F\right)=& \;\bar F_{\bar p}Z,_{\bar
p}-\frac{1}{2}\left(\bar \chi_{\bar p}\bar
\chi_{\bar q}Z,_{\bar p\bar q}\right)\nonumber \\
&+ c\int^1_{-1}d\xi \xi\left[\partial^\mu\bar\phi_{\bar
p}\partial_\mu \bar\phi_{\bar q}K,^\xi_{\bar p\bar q}+
\nabla^2\bar\phi_{\bar p}K^\xi_{\bar p}\right]\nonumber\\
\label{Ygral}
\end{align}
 Now, calling
\be
\label{Lkin} \int d^4y L_{K}\equiv \int d^4y \int d^2\theta
d^2\bar\theta K_*\left(\Phi^i,\bar\Phi^{\bar j}\right)
\ee
it can be shown that
\begin{align}
\label{Lkin1}
L_{K}=&\frac{1}{2}F_iY,_i + \frac{1}{2}\partial^\mu\bar\phi_{\bar p}
\partial_\mu\bar\phi_{\bar q}Z,_{\bar p\bar q}+\frac{1}{2}
\nabla^2\bar\phi_{\bar p}Z,_{\bar p}-\frac{1}{4}\left(\chi^i\chi^j\right)
Y,_{ij}\nonumber\\
&-\frac{1}{2}i\left(\chi^i\sigma^\mu\bar\chi^{\bar p}\right)
\partial_\mu\bar\phi_{\bar q}
Z,_{i\bar p\bar q}-\frac{1}{2}i\left(\chi^i\sigma^\mu
\partial_\mu\bar\chi^{\bar p}\right)Z,_{i\bar p}
\end{align}
%

%%%%%%%%%%%%%%%%%%%%%%%%%%%%%%%%%%%%%%%%%%%%%%%%%%%%%%%%%%%%%%%%%%%%%%%%%%
%%%%%%%%%%%%%%%%%%%%%%%%%%%%%%%%%%%%%%%%%%%%%%%%%%%%%%%%%%%%%%%%%%%%%%%%%%
\subsection{Vacuum properties in deformed theories}

The choice of deforming the anticommutator of $\theta_\alpha$
\eqref{uno}, without altering that of $\bar
\theta_{\dot\alpha}$ implies that $\bar \theta_{\dot\alpha}$
are not the complex conjugate of $\theta_\alpha$, which is only
possible in Euclidean space. Moreover, hermiticity of the
theory is lost because of the deformation and then the usual
analysis of the the potential minima should be replaced by a
careful analysis of the critical points of the resulting
complex expression. At the quantum level, saddle point or
steepest descent methods should be applied as usual, but taking
into account that trajectories are in principle complex and
that space is Euclidean.

 As shown in ref.\cite{Britto} taking the deformed
Wess-Zumino model as a prototype of ${\cal N} =1/2$ theories with
chiral superfields, the vacuum energy, computed from the effective
action for constant bosonic fields, vanishes
\be \langle{\rm vac}| E
|{\rm vac} \rangle = \langle{\rm vac}| Q_\alpha \bar Q_{\dot \alpha}
+ \bar Q_{\dot \alpha} Q_\alpha |{\rm vac} \rangle = 0
\ee
Then,
in order to have a supersymmetric vacuum $Q_\alpha $, the generator
of the surviving supersymmetry, should annihilate both $|{\rm
vac}\rangle$ and $\langle{\rm vac}|$, \be Q_\alpha |{\rm vac}
\rangle = 0  \; , \;\;\; \langle {\rm vac} |Q_\alpha   = 0
\ee
since, being $\bar Q_{\dot\alpha}$  the
generator of the explicitly broken supersymmetry,
 $\bar
Q_{\dot\alpha}|{\rm vac}\rangle$ does not vanish in general.

Vanishing of the vacuum energy for supersymmetric vacua is not
a consequence of any specific choice of the deformed
superpotential. As explained in (\cite{Britto}), supersymmetric
vacua in deformed models with chiral fields impose the
condition $\partial \bar W_* (\bar\Phi)/\partial \bar\Phi = 0$
which in turn imply the vanishing of the corresponding scalar
potential.

%%%%%%%%%%%%%%%%%%%%%%%%%%%%%%%%%%%%%%%%%%%%%%%%%%%%%%%%%%%%%%%%%%%%%%%%%%
%%%%%%%%%%%%%%%%%%%%%%%%%%%%%%%%%%%%%%%%%%%%%%%%%%%%%%%%%%%%%%%%%%%%%%%%%%
%%%%%%%%%%%%%%%%%%%%%%%%%%%%%%%%%%%%%%%%%%%%%%%%%%%%%%%%%%%%%%%%%%%%%%%%%%
%%%%%%%%%%%%%%%%%%%%%%%%%%%%%%%%%%%%%%%%%%%%%%%%%%%%%%%%%%%%%%%%%%%%%%%%%%
\section{Deformed O'Raifeartaigh models}

We discuss here how the landscape of extrema of the scalar
potential in O'Rai\-feartaigh models is affected by the
deformation of superspace defined in eq.(\ref{uno}).

\subsection{Two specific cases}
Consider three chiral superfields fields $\Phi_i$ ($i=1,2,3$)
and a canonical K\"ahler potential $K =\bar \Phi_i * \Phi_i $.
Concerning the superpotential, we choose
\be \label{OR1} {\mathcal{W}} =
\Phi_{1}*\left(\frac{h}{2}\Phi_{3}*\Phi_{3} +
f\right)+m\Phi_{2}*\Phi_{3} + {\rm ST} \ee
which has the typical O'Raifeartaigh potential form, extended
to non(anti)com\-mu\-ta\-tive space. Here ST includes all
necessary symmetrizing terms so that the potential is
symmetrized with respect to the $*$ product. For simplicity, we
take all parameters ($f, m, \ldots$) as real numbers. In order
to compute the scalar potential for component fields $\phi$ we
use eq.(\ref{Vescalar}). In view of the form of the
superpotential, ${\mathcal{W}}\left(\phi_i+\xi cF_i \right)$ as
defined in (\ref{wtilde}) will only have terms with powers
$\xi^n$, $n=0,1,2,3$. Moreover, since integrals with odd powers
in $\xi$ vanish we end with
\begin{align}
{\mathcal{W}}\left(\phi_i+\xi cF_i\right)=\phi_1
\left(\frac{h}{2}\left(\phi_3\right)^2+\xi^2c^2\left(F_3\right)^2+f\right)
+\xi^2 c^2h\phi_3F_3F_1\nonumber
\end{align}
so the diffuse superpotential $\widetilde{{\mathcal{W}}}$
becomes
\begin{align}
\widetilde{{\mathcal{W}}}\left(\phi_i,F_i\right)=2\phi_1
\left(\frac{h}{2}\left(\phi_3\right)^2+f\right)
+\frac{2}{3}c^2\left[\left(F_3\right)^2+h\phi_3F_3F_1\right]\nonumber
\end{align}
leading to a scalar potential
\begin{align}
V_E & = F_1\left(\frac{h}{2}\left(\phi_3\right)^2+f\right)
+ m\left(F_2\phi_3+F_3\phi_2\right)+hF_3\phi_1\phi_3-\frac{\det C}{2}
h F_1\left(F_3\right)^2\label{potencial}
\end{align}
The subscript $E$ indicates that we are dealing with the
Euclidean potential which is minus the Minkowski potential.

Using the equations of motion to replace auxiliary fields $F_i$
and putting all fermion fields to zero we end with
\begin{align}
\label{OR1potential}
V_E =&-\left(\frac{h}{2}\left(\phi_3\right)^2+f\right)\left(\frac{h}{2}
\left({\bar{\phi}_{\bar3}}\right)^2+ f\right) - \left(h\phi_1\phi_3+m\phi_2\right)
\left(h \bar{\phi}_{\bar 1}\bar{\phi}_{\bar3}+m \bar{\phi}_{\bar 2}
\right)\nonumber \\
&- m^2\phi_3{\bar\phi}^{\bar3}+ \frac{h}{2}\det C\left(\frac{h}{2}
\left(\bar\phi_{\bar 3}\right)^2+f\right)
\left(h \bar\phi_{\bar1}\bar\phi_{\bar3}+m \bar\phi_{\bar 2}\right)^2
\end{align}
For $C=0$ we recover the ordinary superspace result with a real
potential provided $\phi^*=\bar\phi$.  For $\det C \neq 0$ the
potential becomes complex not only because the term
proportional to $\det C$ is not accompanied by its complex
conjugate but also because in principle $\bar\phi$ is not the
complex conjugate of $\phi$.

The equations for the extrema of potential (\ref{OR1potential}) read
\begin{align}
\label{OR1extremo1}
0&= h\phi_3 \left(h \bar \phi_{\bar 1}\bar \phi_{\bar 3} +
m \bar\phi_{\bar 2}\right)\\
\label{OR1extremo2}
0&= m\left(h \bar \phi_{\bar 1}\bar
\phi_{\bar 3}+m \bar\phi_{\bar 2}\right)\\
\label{OR1extremo3}
0&= h\phi_3\left(\frac{h}{2}
\left(\bar\phi_{\bar 3}\right)^2+f\right)+m^2
{\bar\phi}_{\bar3}+h\phi_1\left(h \bar \phi_{\bar 1}\bar \phi_{\bar 3}+
m \bar\phi_{\bar 2}\right)\\
\label{OR1extremo4}
0&= \left(h\phi_1\phi_3+
m\phi_2\right)h\bar\phi_{\bar3}\nonumber\\
- &\det Ch\left(\frac{h}{2}\left(\bar\phi_{\bar 3}\right)^2+ f\right)
\left(h \bar\phi_{\bar1}\bar\phi_{\bar3}+m \bar\phi_{\bar 2}
\right)h\bar\phi_{\bar3}\\
\label{OR1extremo5}
0&= m\left(h\phi_1\phi_3 + m\phi_2\right)-\det C hm\left(\frac{h}{2}
\left(\bar\phi_{\bar 3}\right)^2+ f\right)\left(h
\bar\phi_{\bar1}\bar\phi_{\bar3}+m\bar\phi_{\bar2}\right)\\
0&= \left(\frac{h}{2}\left(\phi_3\right)^2+f\right)h\bar\phi_{\bar 3}+
\left(h\phi_1\phi_3+m\phi_2\right)h\bar\phi_{\bar1}+m^2\phi_3\nonumber\\
&- \det C\frac{h}{2}\left(h\bar\phi_{\bar3}\big(h \bar\phi_{\bar1}\bar
\phi_{\bar3}+m \bar\phi_{\bar 2}\big)^2+2\left(\frac{h}{2}
({\bar\phi_{\bar3}})^2 +
f\right)\big(h \bar\phi_{\bar1}\bar\phi_{\bar3}+m \bar\phi_{\bar 2}
\big)h\bar\phi_{\bar1}\right)
\label{OR1extremo6}
\end{align}
Let us first consider the case $m \ne 0$. In this case,
eq.(\ref{OR1extremo2}) implies
\be h \bar \phi_{\bar 1}\bar \phi_{\bar 3}+ m \bar\phi_{\bar 2}=0
\ee
The l.h.s of this equation appears as a factor in all terms
containing $\det C$ and hence all dependence on
$C_{\alpha\beta}$ disappears. Field configurations
corresponding to extrema of the potential are not affected by
the deformation. Moreover, the value of the potential at the
extrema is also unaffected by non(anti)commutativity since
terms containing $\det C$ are multiplied by the same vanishing
factor. The only difference with an ordinary superspace theory
is that, in principle, $\bar \phi_{\bar i}$ does not
necessarily coincide with $\phi_i^*$. For the particular field
configurations where  $\bar \phi_{\bar i}  = \phi_i^*$,  the
results for the undeformed case \cite{Seilectures} apply, and
we can conclude that there is symmetry breaking, no runaway
directions, and a classical pseudomoduli space with degenerate
non supersymmetric vacua (arbitrary $\phi_1^{vac}$).

Concerning the general case in which $\bar \phi_{\bar i}  \ne
\phi_i^*$, we find extrema with  similar properties as those
with $\bar \phi_{\bar i}  = \phi_i^*$ discussed above except
that the pseudomoduli is spanned here by $\phi_1$ and
$\bar\phi_1$ and hence its dimension is doubled. We conclude
the discussion of the  $m \ne 0$ case noting that the theory
above corresponds to a generic supersymmetry breaking potential
because the equation $V=0$ cannot be generically solved.

We will show that the situation changes when the coefficient
$m$ in (\ref{OR1}) vanishes. In that case the $\phi_2$ field
decouples and the scalar potential takes the form
\begin{align}
V_E =&-
\left(\frac{h}{2}\left(\phi_3\right)^2+f\right)\left(\frac{h}{2}
\left({\bar{\phi}_{\bar3}}\right)^2+ f\right) -
h^2\phi_1\phi_3 \bar{\phi}_{\bar 1}\bar{\phi}_{\bar3}
\nonumber\\
& + \frac{h^3}{2}\det C\left(\frac{h}{2}
\left(\bar\phi_{\bar 3}\right)^2+f\right)
\left( \bar\phi_{\bar1}\bar\phi_{\bar3}\right)^2
\end{align}
In the undeformed case we can easily see that there exist two
supersymmetric vacua which correspond to $\phi_1^{vac}=0$ and
$\phi_3 = \pm\sqrt{-2f/h}$ and a supersymmetry breaking flat
direction for $\phi_3^{vac} = 0$, $\phi_1^{vac}$ arbitrary, for
which $V = f^2$ (in Minkowski space).

In the deformed model there are also  six families of
supersymmetric configurations which do not depend on $\det C$.
Namely
\begin{align}
&\bar\phi_{\bar 1}=0\;, &&\bar\phi_{\bar 3}=
\pm \sqrt{-\frac{2f}{h}}\; , \;\;\; \\
&\phi_3=0\;, &&\bar\phi_{\bar 3}=\pm \sqrt{-\frac{2f}{h}}\\
&\bar\phi_{\bar 3}=0\;, &&\phi_3=\pm \sqrt{-\frac{2f}{h}}
\end{align}
All other fields not included in each line are arbitrary.

Concerning non-supersymmetric extrema, they are the same for
the undeformed and the deformed case,
\be \phi_3 = \bar\phi_{\bar 3} = 0 \; , \;\;\; \phi_1 \; \text{
and} \; \bar \phi_{\bar 1} \;\; \text{arbitrary} \ee
and for these configurations   $V_E = -f^2$.

There are also four solutions for which the fields at the
extrema depend on $\det C$
\bea \phi_1 = \phi_3 = 0 \;, \;\;\;\;
\bar\phi_{\bar 1} = \pm \frac{1}{h\sqrt{-\det C}} \;, \;\;\;\;
\bar\phi_{\bar 3} = \pm \sqrt{- \frac{2f}{h}}
\label{OR1 extrdef}
\eea
For these configurations $V=0$ and hence they correspond to
supersymmetric vacua. A remarkable feature of these extrema can
be seen by taking $\det C \in {\mathbb R}$. Indeed, in that
case, in the $\det C\rightarrow0^+$ limit, they correspond to
runaway directions which do not satisfy the extrema conditions
of the undeformed potential. Hence, they have emerged entirely
as a consequence of the deformation.

Let us now consider the vacua structure of another potential
which results from the following superpotential
\be \label{OR2}
{\mathcal{W}}=h\Phi_{1}*\Phi_{3}*\left(\Phi_{3}-m_1\right) +
m\Phi_{2}*\left(\Phi_{3}-m_1\right) + {\rm ST} \ee
In contrast with the superpotential (\ref{OR1}), the form of
this superpotential allows for the existence of critical points
$\partial{\mathcal{W}}/\partial\phi_1 =
\partial{\mathcal{W}}/ \partial\phi_2 = 0$.

A completely analogous calculation to that presented above
leads to the following expression for the scalar potential
\begin{align}
\label{OR2potential}
V_E=&-\left(h\phi_3\left(\phi_3-m_1\right)\right)\left(h\bar\phi_{\bar3}
\left(\bar\phi_{\bar3}-m_1\right)\right)
-m\left(\phi_3-m_1\right)m\left(\bar\phi_{\bar3}-m_1\right)\nonumber\\
-&\left[h\phi_1\left(2\phi_3-m_1\right)+m\phi_2\right]
\left[h\bar\phi_{\bar1}\left(2\bar\phi_{\bar3}-m_1\right)+
m\bar\phi_{\bar2}\right]\nonumber\\
+&\det C\,h\left[h\bar\phi_{\bar1}\left(2\bar\phi_{\bar3}-m_1\right)+
m\bar\phi_{\bar2}\right]^2\left(h\bar\phi_{\bar3}
\left(\bar\phi_{\bar3}-m_1\right)\right)
\end{align}
The equations for the extrema of potential (\ref{OR2potential})
read
\begin{align}
0=&\frac{\partial V}{\partial\phi_1}=
h\left(2\phi_3-m_1\right)\left[h\bar\phi_{\bar1}\left(2\bar\phi_{\bar3}+m_1\right)+
m\bar\phi_{\bar2}\right]
\label{OR2extremo1}\\
\label{OR2extremo2}
0=&\frac{\partial V}{\partial\phi_2}=m
\left[h\bar\phi_{\bar1}\left(2\bar\phi_{\bar3}-m_1\right)+
m\bar\phi_{\bar2}\right]\\
\label{OR2extremo3}
0=&\frac{\partial V}{\partial\phi_3}= (2h\phi_3-hm_1)
\left(h\bar\phi_{\bar3}\left(\bar\phi_{\bar3}-m_1\right)\right)
+m^2\left(\bar\phi_{\bar3}-m_1\right)\\
&+2h\phi_1\left[h\bar\phi_{\bar1}\left(2\bar\phi_{\bar3}-
m_1\right)+m\bar\phi_{\bar2}\right]\nonumber\\
0=&\frac{\partial V}{\partial{\bar\phi_{\bar1}}}=
h\left(2\bar\phi_{\bar3}-m_1\right)\left[h\phi_1\left(2\phi_3-m_1\right)
+m\phi_2\right] \nonumber\\
&- \det C\, 2 h\left[h\bar\phi_{\bar1}\left(2\bar\phi_{\bar3}-m_1\right)+
m\bar\phi_{\bar2}\right]h\left(2\bar\phi_{\bar3}-m_1\right)
\left(h\bar\phi_{\bar3}\left(\bar\phi_{\bar1}-m_1\right)\right)
\label{OR2extremo4}\\
0=&\frac{\partial V}{\partial{\bar\phi_{\bar2}}}=
m\left[h\phi_{1}\left(2\phi_{3}-m_1\right)+m\phi_{2}
\right] \nonumber\\
&- \det C\,2 h m\left[h\bar\phi_{\bar1}
\left(2\bar\phi_{\bar3}-m_1\right)+m\bar\phi_{\bar2}\right]
\left(h\bar\phi_{\bar3}\left(\bar\phi_{\bar3}-m_1\right)\right)
\label{OR2extremo5}
\end{align}
\begin{align}
 \label{OR2extremo6} 0=&\frac{\partial
V}{\partial{\bar\phi_{\bar3}}}=(2h\bar\phi_{\bar3}-hm_1)
\left(h\phi_3\left(\phi_3-m_1\right)\right)
+\left(m\right)^2\left(\phi_3-m_1\right)\\
&+ 2h\bar\phi_1\left[h\phi_1
\left(2\phi_3-m_1\right)+m\phi_2\right]\nonumber\\
&- \det C\, h 2\left[h\bar\phi_{\bar1}\left(2\bar\phi_{\bar3}-m_1\right)+
m\bar\phi_{\bar2}\right]2h\bar\phi_1h\bar\phi_{\bar3}
\left(\bar\phi_{\bar3}-m_1\right)\nonumber\\
&- \det C\, h\left[h\bar\phi_{\bar1}\left(2\bar\phi_{\bar3}-m_1
\right)+m\bar\phi_{\bar2}\right]^2\left(2h\bar\phi_{\bar3}-hm_1\right)\nonumber
\end{align}
As in the previous example, let us first consider the case $m
\ne 0$. In that case, eq.(\ref{OR2extremo2}) implies
\be \label{m2noCero}
h\bar\phi_{\bar1}\left(2\bar\phi_{\bar3}-m_1\right)+
m\bar\phi_{\bar2}=0 \ee
Again, the l.h.s of this equation appears as a factor in all
terms containing $\det C$ and hence all dependence on
$C_{\alpha\beta}$ disappears and field configurations
corresponding to extrema of the potential are not affected by
the deformation. Moreover, the value of the potential is also
unaffected by non(anti)commutativity since terms containing
$\det C$ are multiplied by the same vanishing factor. As
explained in \cite{Seilectures} there are supersymmetric vacua
$\phi_i^S$ which corresponds to
\be
\phi_3^S  = m_1 \; , \;\;\;  \phi_2^S = -\frac{hm_1}{m} \phi_1^S
\ee
(in the deformed case we should have identical values for
fields $\bar \phi_{\bar i}$ which, in the deformed case are not
automatically related to $\phi_i$).

As in the undeformed case, there are also extrema $\phi^M$  for
which $V[\phi^M ] \ne 0$. In fact, the Euclidean  $V[\phi^M]$
is a real negative number which in Minkowski undeformed
superspace would lead to the metastable vacua. The explicit
form of the solutions is the same as in the undeformed case.

Let us now consider the $m=0$ case. In the undeformed
(Minkowski) space, the non-supersymmetric (metastable) vacua
present for $m \ne 0$ are lost but, as we will see, the
situation changes in the deformed case. Indeed for vanishing
$m$ the scalar potential takes the form
\begin{align} \label{OR2potentialm2Nulo} V_E
=&-h\phi_3\left(\phi_3-m_1\right)
h\bar\phi_{\bar3}\left(\bar\phi_{\bar3}-m_1\right) \nonumber\\
&- h\phi_1\left(2\phi_3-m_1\right)
h\bar\phi_{\bar1}\left(2\bar\phi_{\bar3}-m_1\right) \nonumber\\
&+\det Ch^2\left[h\bar\phi_{\bar1}
\left(2\bar\phi_{\bar3}-m_1\right)\right]^2
\bar\phi_{\bar3}\left(\bar\phi_{\bar3}-m_1\right)
\end{align}
Let us compare the supersymmetric vacuum states between the
undeformed and the deformed case. In the undeformed case, we
have four supersymmetric vacuum states:
\begin{align} \label{vacuaOR2}
&\phi_1=\bar\phi_{\bar 1} = \phi_3=\bar\phi_{\bar 3}=0\nonumber\\
&\phi_1=\bar\phi_{\bar 1} = \phi_3=0\;, \;\;\bar\phi_{\bar 3} = m_1\nonumber\\
&\phi_1=\bar\phi_{\bar 1} = \bar\phi_{\bar 3}=0\;,\;\; \phi_3=m_1\nonumber\\
&\phi_1=\bar\phi_{\bar 1} = 0\;,\;\;\phi_3=\bar\phi_{\bar 3} = m_1
\end{align}
In the deformed case, the vacua (\ref{vacuaOR2}) are still
present. In addition, there are other four supersymmetric
vacua:
\begin{align}
\label{OR2ExtrDep de detC}
&\phi_1=\bar\phi_{\bar 3} = 0\;,\;\; \bar\phi_{\bar 1}=
\pm\frac{i}{2h\sqrt{-\det C}}\;,\;\;\phi_3=m_1/2 \nonumber\\
&\phi_1=\bar\phi_{\bar 3} = 0\;,\;\; \bar\phi_{\bar 1}=
\pm\frac{i}{2h\sqrt{-\det C}}\;,\;\;\phi_3=m_1
\end{align}
As in the case of the extrema (\ref{OR1 extrdef}) of the
previous example, in the limit $\det C\rightarrow0^+$ these
extrema correspond to runaway directions which do not exist in
the case of the undeformed potential $\det C = 0$.

Concerning the supersymmetry breaking vacua, there is no difference
between the undeformed and deformed case, having in both the
pseudomoduli space:
\begin{align} \label{OR2 meta} \phi_3=\bar\phi_3=m_1/2
\end{align}
for which $V=(m_1/2)^4h^2$ in the undeformed case and
$V_E=-(m_1/2)^4h^2$ in the deformed one.

\subsection{A more general superpotential}

We end this section discussing conditions on a general cubic
superpotential under which the vacuum structure remains
unaffected by the deformation. Consider $n$ chiral superfields
$\Phi_i$ ($i=1,2,\ldots,n$), a canonical K\"ahler potential and
a deformed superpotential of the form
\be \label{Superpotencialgeneral} {\mathcal{W}}
\left(\Phi_p\right)=
C+C_{q}\Phi_{q}+C_{qr}\Phi_q*\Phi_r+C_{qrs}\Phi_q*\Phi_r*\Phi_s
\ee
with $C, C_q, C_{qr},$ y $C_{qrs}$ arbitrary coefficients,
symmetric in all their indices. As before, in view of the form
of the superpotential, the functional
${\mathcal{W}}\left(\phi_i+\xi cF_i \right)$, as defined in
(\ref{wtilde}), will just contain terms with powers 0, 1, 2,
and 3 of $\xi$. Only even powers will contribute to
$\tilde{\mathcal{W}}$ obtaining
\begin{align}
{\mathcal{W}}\left(\phi_i+\xi cF_i \right)=&
C+C_{q}\phi_q+C_{qr} \phi_q\phi_r+C_{qrs}\phi_q\phi_r\phi_s
\nonumber\\
&+\xi^2c^2\big( C_{qr}F_qF_r+3C_{qrs}\left(\phi_qF_rF_s\right)\big)\\
\widetilde{{\mathcal{W}}}\big(\phi_i,F_i\big)=&
2\left(C+C_{q}\phi_q+C_{qr}\phi_q\phi_r+C_{qrs}\phi_q\phi_r\phi_s\right)
\nonumber\\
&+\frac{2c^2}{3}\big(C_{qr}F_qF_r+3C_{qrs}\left(\phi_qF_rF_s\right)\big)
\end{align}
Using the equations of motion for auxiliary fields $\bar
F_{\bar j}$ we find  $F_i=-{\partial \bar
{\mathcal{W}}}/{\partial\bar\phi_i}$ and then
\be {\widetilde{{\mathcal{W}}}},_i=2\left(C_i+2C_{ir}\phi_r +
3C_{irs}\phi_r\phi_s\right)+ 2c^2C_{irs}F_rF_s \ee
With this \be \label{PotencialGeneral}
V=\left.F_i\left[C_i+2C_{ir}\phi_r+3C_{irs}\phi_r\phi_s +
c^2C_{irs}F_rF_s\right]
\right\vert_{F_i=F_i\left(\phi_j,\bar\phi_{\bar j}\right)} \ee
The extrema conditions are
\begin{align}
\label{extremosORgeneral}
0&=\frac{\partial V}{\partial\phi_j}=2\left.F_i\left[C_{ij}+
3C_{ijr}\phi_r\right]\right
\vert_{F_i=F_i\left(\phi_j,\bar\phi_{\bar j}\right)}
\\
\label{extremosORgeneralTecho}
0 &=  \frac{\partial V}{\partial\bar\phi_{\bar j}}= \left.F_i,_{\bar j}
\left[C_i+2C_{ir}\phi_r+3C_{irs}\phi_r\phi_s+3\, c^2C_{irs}F_rF_s\right]
\right\vert_{F_i=F_i\left(\phi_j,\bar\phi_{\bar j}\right)}\nonumber\\
\end{align}

Suppose  that the following relations among coefficients
$C_{ij}$ and $C_{ijr}$ hold
\be \label{CondicionParaM0} \big(C_{ij}+3C_{ijr}\phi_r\big) =
\delta_{ia}\,M_j + \delta_{ja} \, M_i \ee
for some value $a$ ($M_i$ is an arbitrary, field dependent,
vector). Such conditions imply that $F_a=0$ (unless, for all
$i$, the pairs of coefficients $(C_i, C_{iaa})$ are
proportional to each other, cf. (\ref{extremosORgeneral})). If
we still impose a more restrictive condition on $C_{ijr}$,
namely that it vanishes unless it has two indices $a$, we see
that the extrema conditions (\ref{extremosORgeneralTecho}) are
independent of $\det C$ and also the potential at the extrema
is unaffected by the deformation.

Is easy to see that the above mentioned conditions force the
potential to take the form
\be \label{ORgeneral} \mathcal{W}=\sum_{i\neq a}
\Phi_i * g_i(\Phi_a) + \text{ST}\ee
with  $g_i$ quadratic functions, not all proportional to each other.

By the above arguments, the vacuum structure  of this
superpotential is not deformed. Note that the explicit examples
previously discussed in subsection 3.2 belong (for $m \ne 0$) to
this class of potentials, insensitive to the deformations.

%%%%%%%%%%%%%%%%%%%%%%%%%%%%%%%%%%%%%%%%%%%%%%%%%%%%%%%%%%%%%%%%%%%%%%%%%%
%%%%%%%%%%%%%%%%%%%%%%%%%%%%%%%%%%%%%%%%%%%%%%%%%%%%%%%%%%%%%%%%%%%%%%%%%%
%%%%%%%%%%%%%%%%%%%%%%%%%%%%%%%%%%%%%%%%%%%%%%%%%%%%%%%%%%%%%%%%%%%%%%%%%%
%%%%%%%%%%%%%%%%%%%%%%%%%%%%%%%%%%%%%%%%%%%%%%%%%%%%%%%%%%%%%%%%%%%%%%%%%%
\section{Noncanonical deformed K\"ahler potentials}

As a first simple example of noncanonical K\"ahler potential we
consider
\be K=\left(\Phi* \bar \Phi\right)^2 \label{cuartico} \ee
In this case eqs. (\ref{Zgral}) and
(\ref{Ygral}) take the form
\be Z\left(\phi,\bar\phi,F\right)=\int^1_{-1}d\xi
\left(\phi\bar\phi+\xi cF\bar\phi\right)^2=
2\left(\phi\bar\phi\right)^2+ \frac{2}{3}c^2\left(\bar\phi
F\right)^2\nonumber
\ee
\begin{align}
Y\left(\phi,\bar\phi,F,\bar F\right)=&\bar F\left(4\phi^2\bar\phi+
\frac{4}{3}c^2F^2\bar\phi\right)-
\frac{1}{2}\left(\bar\chi\bar\chi\right)
\left(4\phi^2+\frac{4}{3}c^2F^2\right)\nonumber\\
&+ c\int_{-1}^1d\xi\xi \left[\partial^\mu\bar\phi\partial_\mu\bar\phi
\left(2\phi^2+4\xi c\phi F+2\xi^2c^2F^2\right)\right.\nonumber\\
&+ 2\Box\bar\phi\left.\left(
\phi\bar\phi+\xi cF\bar\phi\right)\left(\phi+\xi cF\right)\right]\nonumber\\
=& \bar F\left(4\phi^2\bar\phi+\frac{4}{3}c^2F^2\bar\phi\right)-
\frac{1}{2}\left(\bar\chi\bar\chi\right)\left(4\phi^2
+\frac{4}{3}c^2F^2\bar\phi\right)\nonumber\\
&+ \left(\frac{4}{3}c^2\phi F\right)
\left[2\partial^\mu\bar\phi\partial_\mu\bar\phi +
\bar\phi\Box\bar\phi\right]\nonumber
\end{align}
so that the kinetic part of the component field Lagrangian
reads
\begin{align} \label{LkinNOcan}
L_{K}=&  -4\bar F F \phi\bar\phi +2F \left(\bar\chi\bar\chi\right)\phi
+\frac{2}{3}{\det C}  F^2\left[2\partial^\mu\bar\phi\partial_\mu\bar\phi
+\bar\phi\Box\bar\phi\right] \\
&-\frac{1}{2}\partial^\mu\bar\phi\partial_\mu\bar\phi\left(4\phi^2-
\frac{4}{3} \det C F^2\right)
-\frac{1}{2}\Box\bar\phi\left(4\phi^2\bar\phi+
\frac{4}{3}c^2F^2\bar\phi\right)\nonumber\\
&-\frac{1}{4}\chi\chi\left(8\bar F\bar\phi-4\bar\chi\bar\chi\right)
+\frac{1}{2}i\left(\chi\sigma^\mu\bar\chi\right)
\partial_\mu\bar\phi8\phi+\frac{1}{2}i\left(\chi\sigma^\mu
\partial_\mu\bar\chi\right)
8\phi\bar\phi \nonumber\\
\end{align}
Since $\det C$ only affects kinetic energy terms for $\bar
\phi$, the scalar potential for this noncanonical K\"ahler
potential could only be deformed by contributions arising from
the superpotential.

Because $L_{K}$ is a linear functional of the K\"ahler
potential the discussion above also applies to the case
\be \label{kahler X aprox0} K=\Phi * \bar
\Phi+\lambda\left(\Phi* {\bar \Phi}\right)^2 \ee
Such a K\"ahler potential can be though as resulting from the
approximation of a general potential
$K_{*}\left(\Phi,\bar\Phi\right)=f\left(\Phi * \bar
\Phi\right)$ for $\Phi\approx 0$. Then, in the weak-field
regime we have to expect that only the deformation of the
superpotential would affect the vacuum structure.

Modifications arise for K\"ahler potentials with higher powers,
namely $\left(\bar\Phi\Phi\right)^n$ with $n>2$. Consider the
simplest case $n=3$,
\be \label{sextico}
K_3 = \left(\bar\Phi * \Phi\right)^3
\ee
Since we are interested in purely bosonic contributions with no
derivatives, we will restrict our analysis to these type of
terms which will be indicated with the subscript ``boson". We
have,
\be Z\left(\phi,\bar\phi,F\right)=
\int^1_{-1}d\xi\left(\phi\bar\phi+\xi cF\bar\phi\right)^3=
2\left[\left(\phi\bar\phi\right)^3- \det C\bar\phi^3\phi
F^2\right]\nonumber \ee
\begin{align}
Y_\text{boson}\left(\phi,\bar\phi,F,\bar F\right) &= 6\bar F
\left[\phi^3\bar\phi^2-
\det C\bar\phi^2\phi F^2\right]\nonumber\\
\frac{\partial Y_\text{boson}}{\partial\phi} &= 6\bar F
\left[3\phi^2\bar\phi^2-
\det C\bar\phi^2F^2\right]\nonumber
\end{align}
The corresponding contribution to the Lagrangian is,
\begin{align}
\label{LkinNOcan2}
\left. L_{K_3}\right|_\text{boson}
= 3 F\bar F\bar\phi^2 \left(3\phi^2 - \det C F^2\right)
\end{align}
so the relevant parts of the equations of motion for the
auxiliary fields are
\begin{align}
3F\bar\phi^2\left(3\phi^2-\det CF^2\right)+
\frac{\partial \bar {\mathcal{W}}}{\partial\bar \phi}&= 0\label{esta2}\\
9\bar F\bar\phi^2\phi^2-9\det C F^2\bar F\bar\phi^2 +
\frac{\partial {\mathcal{W}}}{\partial \phi}&= 0
\label{aquella2}
\end{align}
We then conclude that both $F$ and $\bar F$ will depend on
$\det C$ independently of the choice of the superpotential, so
that for a K\"ahler potential cubic in $\bar\Phi * \Phi$ the
scalar potential and, a fortiori, the vacuum structure will be
affected by the deformation.

Let us consider a simple example that illustrates the discussion
above. It corresponds to  superpotentials  ${\cal W}$ and ${\cal
\bar W}$ (recall that in Euclidean space, they are independent
functionals)
\bea
    {\cal W}=\frac12f\Phi*\Phi\;\;\;,\;\;\;{\cal \bar W}=g\label{pseudo}
\eea
and the K\"ahler potential defined in (\ref{sextico}).

Given superpotentials (\ref{pseudo}) we get for the auxiliary
fields, using eqs. of motion (\ref{esta2}) and
(\ref{aquella2}),
\bea
    F&=&\frac{i\sqrt3 \phi}{\det C} \label{f1}\\
    \bar F&=& \frac{f}{36\phi\bar\phi^2}\label{f2}
\eea
It can be seen from eq.(\ref{Vescalar}) that, as expected, the
scalar potential is affected by the deformation of the K\"ahler
potential through the dependence of $F$ on $\det C$ as given by
(\ref{f1}).

Let us end this section by pointing that a completely analogous
behavior can be found for a K\"ahler potential of the form $K_n
= (\bar\Phi*\Phi)^n$. For example, for odd $n$ we find, instead
of eq.(\ref{esta2}), that the auxiliary field $F$ obeys the
equation
\be \frac{n}{2 \sqrt{-\det C}}\,
\bar\phi^{n-1}\left((\phi+\sqrt{-\det C} F)^n-(\phi-\sqrt{-\det
C}F)^n\right)+ \frac{\partial\bar W}{\partial\bar\phi}=0 \ee
This is a degree $n$ polynomial equation for $F$,  with
coefficients depending on $\det C$ as a result of the
deformation in the K\"ahler potential. The solution for $F$
will be in general $\det C$-dependent (as we have explicitly
seen for the particular case $n=3$) and hence the scalar
potential as given by (\ref{Vescalar}) will in turn be
deformed.

%%%%%%%%%%%%%%%%%%%%%%%%%%%%%%%%%%%%%%%%%%%%%%%%%%%%%%%%%%%%%%%%%%%%%%%%%%
%%%%%%%%%%%%%%%%%%%%%%%%%%%%%%%%%%%%%%%%%%%%%%%%%%%%%%%%%%%%%%%%%%%%%%%%%%
%%%%%%%%%%%%%%%%%%%%%%%%%%%%%%%%%%%%%%%%%%%%%%%%%%%%%%%%%%%%%%%%%%%%%%%%%%
%%%%%%%%%%%%%%%%%%%%%%%%%%%%%%%%%%%%%%%%%%%%%%%%%%%%%%%%%%%%%%%%%%%%%%%%%%
\section{Discussion}

In this work we have discussed the vacuum structure of ${\cal
N} = 1/2$ supersymmetric theories of chiral superfields in
deformed superspace. We have analyzed O'Raifeartaigh models
with general deformed superpotentials, including the case in
which the K\"ahler potential is non-canonical. The question we
intended to clarify was how the landscape of extrema of the
classical scalar potential is affected by a deformation of
superspace.

As explained in section 2.3, although hermiticity of the theory
is lost because of the deformation, the analysis of the
critical points of the resulting complex potential allows to
decide whether the ${\cal N} = 1/2$ supersymmetry surviving the
deformation is spontaneously unbroken. In fact, as we have
seen, loss of hermiticity implies that the scalar potential is
in principle complex and, moreover, because superfields $\bar
\Phi_i$ are not the complex conjugate of $\Phi_i$, scalars
$\bar \phi_{\bar i}$ do not in general coincide with
$\phi_i^*$.  This of course complicates the analysis of extrema
of the potential unless we impose some restrictions on fields
and potentials.

Restricting the analysis to the case of field configurations
such that $\bar \phi_{\bar i}  = \phi_i^*$,  we have seen in
section 3.2 that the vacuum configurations for  superpotentials
(\ref{OR1}) and (\ref{OR2}) described in
\cite{Sei0}-\cite{Seilectures} for undeformed superspace, are
also present in the deformed case when the coefficient $m \ne
0$. Hence  in both cases there is symmetry breaking and a
classical pseudomoduli space with degenerate non supersymmetric
vacua. The difference between the two cases is that in the
latter there can be  metastable (for an appropriate choice of
coefficients) vacua which are absent in the former.

An interesting phenomenon takes place for $m=0$: in the limit
$\det C\rightarrow 0$, in which the deformation vanishes, there
are additional extrema, eqs. (\ref{OR1 extrdef}) and
(\ref{OR2ExtrDep de detC}), that correspond to runaway
directions which do not exist in the case of the undeformed
potential $\det C = 0$. This phenomenon is resemblant of what
happens with solitons in $\theta$-deformed noncommutative
space: apart from those that reproduce the ordinary regular
solitons in the $\theta \to 0$ limit, there are ``fluxon''
solutions with no regular counterpart in ordinary space (see
\cite{FAS} and reference therein).

In section 3.3  we considered  a general cubic superpotential
(which encompasses the two previous examples) and found the
conditions under which the vacuum structure remains unaffected
by the deformation.

We also considered non-canonical deformed K\"ahler potentials
which, being non-quadratic, could be expected to induce a
$C$-dependence on the vacuum structure. The case $K = (\bar\Phi
*\Phi)^2$ is a counterexample of this possibility since we
proved that only the kinetic energy is affected by the
deformation. Hence, in a weak-field approximation, the vacuum
dependence on the $C$-deformation will only enter through the
deformed superpotential. We need higher powers ($n>2$) of $\bar
\Phi* \Phi$ in order to change the vacuum structure as we have
explicitly shown at the end of section 4.

The discussion in this work is valid at tree-level, and should
be corrected by including leading quantum corrections to the
potential. Being the theory non-hermitian, one should resort to
complex saddle point or steepest descent methods. We hope to
report on this issue in a following investigation.
O'Raifeartaigh-type  models, as those considered here, can
arise naturally and dynamically in the low-energy limit of
simple SUSY gauge theories. In this respect, the extension of
the analysis we have presented to the case of deformed super
Yang-Mills theory is also a subject we hope to address in the
future.

~

\noindent\underline{Acknowledgements}:   This work was partially
supported by   PIP6160-CONICET,  BID 1728OC/AR PICT20204-ANPCYT
grants and by CIC and UNLP, Argentina.

%%%%%%%%%%%%%%%%%%%%%%%%%%%%%%%%%%%%%%%%%%%%%%%%%%%%%%%%%%%%%%%%%%%%%%%%%%
%%%%%%%%%%%%%%%%%%%%%%%%%%%%%%%%%%%%%%%%%%%%%%%%%%%%%%%%%%%%%%%%%%%%%%%%%%
%%%%%%%%%%%%%%%%%%%%%%%%%%%%%%%%%%%%%%%%%%%%%%%%%%%%%%%%%%%%%%%%%%%%%%%%%%
%%%%%%%%%%%%%%%%%%%%%%%%%%%%%%%%%%%%%%%%%%%%%%%%%%%%%%%%%%%%%%%%%%%%%%%%%%
%%%%%%%%%%%%%%%%%%%%%%%%%%%%%%%%%%%%%%%%%%%%%%%%%%%%%%%%%%%%%%%%%%%%%%%%%%
%%%%%%%%%%%%%%%%%%%%%%%%%%%%%%%%%%%%%%%%%%%%%%%%%%%%%%%%%%%%%%%%%%%%%%%%%%
%%%%%%%%%%%%%%%%%%%%%%%%%%%%%%%%%%%%%%%%%%%%%%%%%%%%%%%%%%%%%%%%%%%%%%%%%%
%%%%%%%%%%%%%%%%%%%%%%%%%%%%%%%%%%%%%%%%%%%%%%%%%%%%%%%%%%%%%%%%%%%%%%%%%%

%
%%%%%%%%%%%%%%%%%%%%%%%%%%%%%%%%%%%%%%%%%%%%%%%%%%%%%%%%%%%%%%%%%%%%%%%%%%
%%%%%%%%%%%%%%%%%%%%%%%%%%%%%%%%%%%%%%%%%%%%%%%%%%%%%%%%%%%%%%%%%%%%%%%%%%
%%%%%%%%%%%%%%%%%%%%%%%%%%%%%%%%%%%%%%%%%%%%%%%%%%%%%%%%%%%%%%%%%%%%%%%%%%
%%%%%%%%%%%%%%%%%%%%%%%%%%%%%%%%%%%%%%%%%%%%%%%%%%%%%%%%%%%%%%%%%%%%%%%%%%
%%%%%%%%%%%%%%%%%%%%%%%%%%%%%%%%%%%%%%%%%%%%%%%%%%%%%%%%%%%%%%%%%%%%%%%%%%

\end{document}